# Exploration of cavitation-induced erosion metrics in throttle flow simulations

[1]Gina M. Magnotti*; [2]Michele Battistoni; [1,3]Kaushik Saha; [1]Sibendu Som

[1]Argonne National Laboratory, Argonne, IL, USA; [2]Università degli Studi di Perugia, Perugia, Italy; [3]Bennett University, Greater Noida, India

**Abstract**

Although there have been extensive experimental and computational investigations in characterizing cavitation phenomenon in both diesel and gasoline direct injectors, much is still unknown about the mechanisms driving cavitation-induced erosion, and how this complicated fluid-structure interaction should be modeled. To explore current modeling capabilities, a numerical investigation was conducted within the CONVERGE modeling framework to assess proposed cavitation erosion metrics in the literature, and their link to the predicted cavitation cloud collapse mechanism. The multiphase flow within the Winklhofer Throttle "U" geometry was modeled using a compressible mixture model, where phase change was represented using the Homogeneous Relaxation Model (HRM) and the turbulent flow evolution was modeled using a dynamic structure approach for Large Eddy Simulations (LES). After comparing the model predictions against available experimental data, representative condensation events and potential cavitation erosion sites were identified. The cavitation cloud structures responsible for potential material damage were visualized through the evolution of the vorticity field. For the modeled throttle geometry, it was found that the horseshoe cloud implosion mechanism was predicted to occur and generate excessive impact loads at the throttle boundary.

**Introduction**

Cavitation-induced erosion has been extensively studied using experimental [1,2,3,4] and computational modeling techniques [4,5,6,7,8,9] to gain insight into the physical mechanisms driving this process for a range of applications, from pump impellers to ship rudders. Cavitation erosion has been noted in diesel injector hardware, namely within the needle seat region and along the injector needle, as well as the entrance to nozzle holes [4,6]. A common feature among these vulnerable locations is the proximity to an area contraction, where local flow acceleration and pressure reduction promote cavitation. The erosive potential of the cavitation shedding processes in these regions have been found to be highly dependent on the local geometric features and fluid properties.

Ideally, an injector hardware designer would be able to utilize a computationally efficient design tool to accurately predict these cavitation and condensation events, and to inform improved designs that mitigate the propensity and severity of cavitation-induced erosion. However, the main challenge of developing such a tool is linking the thermofluidic phenomenon of vapor generation and collapse with material damage to neighboring surfaces. In lieu of computationally expensive fluid-structure interaction modeling [9], the Eulerian mixture modeling approach has been accepted as a computationally efficient means of capturing cavitation phenomena when sufficient resolution is employed [8]. However, when employing such a framework, there remains a need to link condensation events with cavitation erosion potential. Although several cavitation erosion indicators have been proposed in the literature [5,6,7], no single metric has been identified as universally applicable across all injector geometries and injection conditions.

The overarching goal of this work is to identify cavitation cloud collapse mechanisms that are likely to occur within injector orifices, and develop parameters that adequately characterize the cavitation erosion potential of these fluid structures. In order to progress towards identifying a predictive and robust indicator, we focus our efforts on modeling the flow of pressurized fuel through a simplified throttle geometry with an inlet diameter of approximately 300 μm [10], which serves as a geometric analogue to an injector nozzle. We study the cavitation shedding process in detail, and identify cavitation cloud structures that are formed. In the initial stage of this effort, we assess the ability of the maximum predicted pressure at the throttle surface boundaries to characterize the propensity for cavitation erosion [6], and evaluate its link to the observed cavitation structures.

**Model Formulation**

In this work, we utilize CONVERGE [11] to model cavitation within the Winklhofer Throttle "U" geometry [10] for pressure drops ranging from 20 to 85 bar. Although no experimental data exist to characterize erosion events within this geometry, this test geometry serves as the starting point to validate the current cavitation modeling configuration, and study cavitation cloud structures that are predicted. The key parameters describing the experimental conditions explored in this work are defined in Table 1. Diesel fuel is represented using properties for tetradecane, and trace amounts of non-condensable gas, namely $N_2$ comprising a mass fraction of 5.0e-6, are assumed to be present in the liquid fuel. Liquid and gas phases are treated as compressible. Cavitation and condensation are represented using the homogeneous relaxation model (HRM) [12], where the rate at which the instantaneous mass fraction of vapor, $x$, approaches its equilibrium value, $\bar{x}$, is defined with the following relation,

$$\frac{Dx}{Dt} = \frac{\bar{x} - x}{\theta} \qquad (1)$$

where $\theta$ is the relaxation time scale. In this study, cavitation and condensation are assumed to occur at equal timescales, and are calculated as $\theta = \theta_0 \alpha^{-0.54} \psi^{1.76}$, where $\theta_0$ is a coefficient set to 3.84e-7, $\alpha$ is the total void fraction, including fuel vapor and non-condensable gases, and $\psi$ is the non-dimensional pressure ratio ($\psi = \frac{p_{sat}-p}{p_{crit}-p_{sat}}$). To adequately represent the turbulent flow structures within the throttle, the large eddy simulation (LES) dynamic structure model [13] is employed with a minimum grid size of 5.0 µm.

*Corresponding Author, Gina Magnotti: gmagnotti@anl.gov

| Liquid Fuel | Fuel Temperature [K] | Upstream Reservoir Pressure [bar] | Downstream Reservoir Pressure [bar] | Pressure Drop Across Throttle [bar] | Nitrogen Mass Fraction |
|---|---|---|---|---|---|
| Tetradecane | 300 | 100 | 15 - 80 | 20 - 85 | 5.0e-6 |

Table 1. Modeled conditions within the Winklhofer Throttle "U" Geometry

**Results and Discussion**

To validate the model set-up, predictions of mass flow rate at the throttle exit are compared with the experimental data from Winklhofer et al. [10] for a range of pressure drop conditions from 20 to 85 bar, as shown in Figure 1. Overall, excellent agreement is achieved between the predicted and measured mass flow rates across the wide range of pressure drops evaluated in this study. At pressure drop conditions greater than 70 bar where choking of the flow is observed, the level of disagreement between the model predictions and experimental data increases slightly, but remains within 3% of the measured mass flow rate. This validation exercise suggests that the fluid properties and treatment of the multiphase flow in this modeling approach well represent the experimental conditions at the throttle exit.

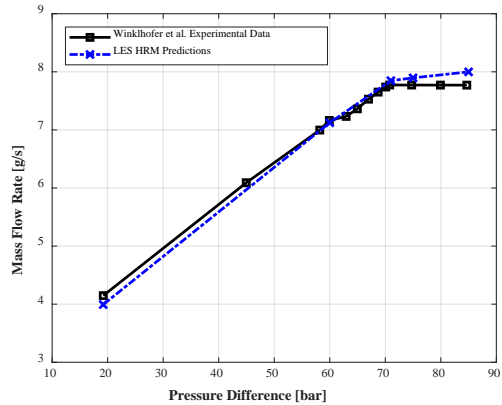

Figure 1. Comparison of measured and predicted mass flow rates for a range of pressure drop conditions across the Winklhofer Throttle "U" geometry [10].

In order to assess the ability of the modeling framework to capture the flow conditions within the throttle, model predictions of the vapor mass fraction distribution within the central plane are compared to the extent of cavitation as indicated in the experimental images of Winklhofer and co-workers [10]. A representative comparison is shown in Figures 2(a) and (b) for a pressure drop condition of 75 bar, where both the experimental and computational images have been time-averaged after steady state conditions have been reached. Cavitation inception is predicted at the inlet corners of the throttle, and extends along the wall boundary until approximately half of the length of the throttle. In comparison to the experimental results in Figure 2(a), less cavitation is predicted. Due to the sensitivity of cavitation inception and development to the inlet geometric features, it is possible that this discrepancy may be due to slight differences between the experimental and simulated geometries. Because the predicted mass flow rate at the throttle exit is within 2% of the experimental measured condition, this result also suggests that the initial flow development at the throttle entrance may not be adequately resolved to capture the flow acceleration and local pressure reduction required to initiate sufficient cavitation formation observed in the experiment. Future work will explore the influence of turbulence model choice and grid resolution on the predicted cavitation structures.

*Corresponding Author, Gina Magnotti: gmagnotti@anl.gov

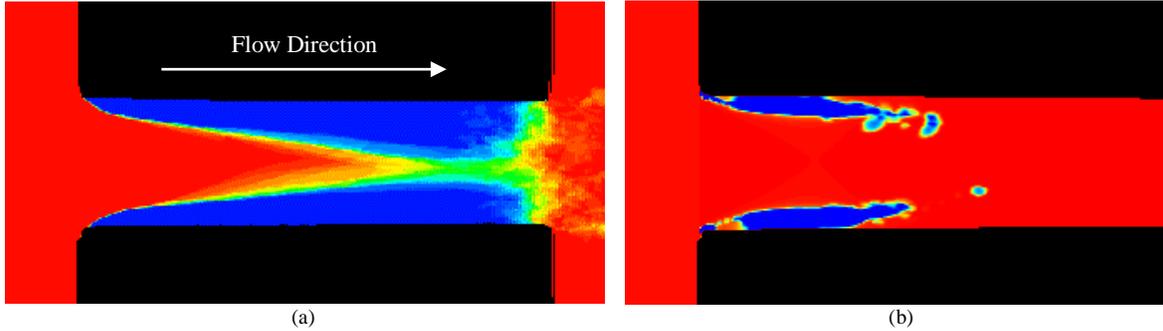

Figure 2. Comparison of total void fraction between (a) experimental image from Winklhofer et al. [10] and (b) time-averaged HRM model prediction at the central slice for a pressure drop of 75 bar.

Although model predictions indicate less cavitation than was experimentally observed, the current results present an opportunity to examine the cavitation shedding process and erosive potential of predicted condensation events. In this computational exploration, the maximum predicted pressure at the throttle boundary, as proposed by Koukouvinis and co-workers [6], is implemented as an indicator of cavitation erosion. Such an indicator aids in highlighting key cavitation cloud collapse structures among the multitude of time steps. A representative cavitation cloud collapse event is shown in Figure 3, where cavitation structures are visualized using iso-contours of 1% total void fraction.

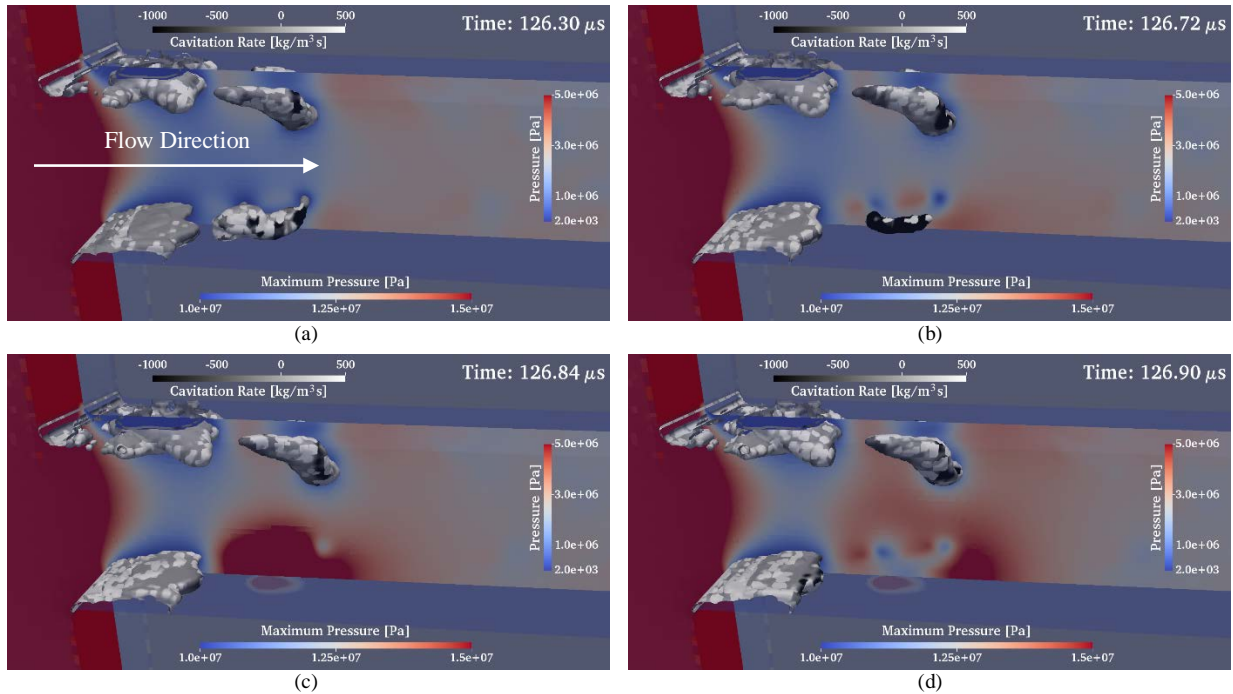

Figure 3. The temporal evolution of cavitation cloud collapse leading to substantial impact loading on the lower throttle boundary is visualized in panels (a) through (d) across a 0.60 µs time period. The boundaries are colored by maximum pressure, the central slice is colored by local pressure, and iso-contours of void fraction are colored by cavitation rate, where negative values indicate condensation.

In the time instant shown in Figure 3(a), fixed cavities form at the entrance to the throttle. The growth and breakup of this cavity lead to the formation of traveling hairpin or horseshoe-shaped cavitation structures. As these structures travel downstream, as shown in Figure 3(b), the pressure recovery region promotes condensation and vapor cloud collapse, as indicated by the dark colored void fraction iso-contours. The rapid vapor cloud collapse induces a strong pressure wave as illustrated in Figure 3(c), and an impact load in excess of 15 MPa is predicted at the neighboring throttle boundary. Although the maximum predicted pressure is below the yield strength of typical injector materials, such as steel with a yield strength greater than 200 MPa [14], it is very likely the acoustic pressure wave could generate higher erosive potential via the collapse of micro-scale cavitation structures or bubbles in the vicinity of the throttle boundary [15,16]. Following this event, a pressure wave is observed to propagate downstream, as shown in Figure 3(d).

*Corresponding Author, Gina Magnotti: gmagnotti@anl.gov

Although the implemented index helps to mark cavitation cloud collapse events that could lead to cavitation erosion, it is not able to explain the physical mechanism governing the cavitation erosion event. In order to better understand the underlying physics, the generation and development of vortices in the cavitation shedding process was analyzed. This investigation is motivated by the work of Dular and Petkovšek [1], where the link between vortex structures and cavitation erosion potential was noted to be of critical importance for all five cavitation erosion mechanisms. The vortex structures, shown in Figure 4, are visualized using iso-contours of the magnitude of the vorticity field. Although this criterion highlights both rotational and shear structures within the flow, the development of the horseshoe vortex can be clearly visualized with appropriate selection of the vorticity magnitude threshold and filtering of the void fraction distribution. These results provide strong evidence that the predicted cavitation erosion event, indicated by the maximum pressure index, was due to the collapse of a horseshoe cloud. Although spherical cloud collapse was also observed in the simulation, the predicted maximum pressure at the wall boundaries was smaller than that from the horseshoe cloud implosion event. Future studies will implement more rigorous criteria for characterizing vortex structures in compressible flows [17] in order to identify and gather statistics regarding the occurrence of other predicted cavitation cloud collapse mechanisms. This exercise will help inform cavitation erosion metrics that are capable of linking the occurrence of cavitation cloud structures with their potential for material damage under injector relevant flow conditions.

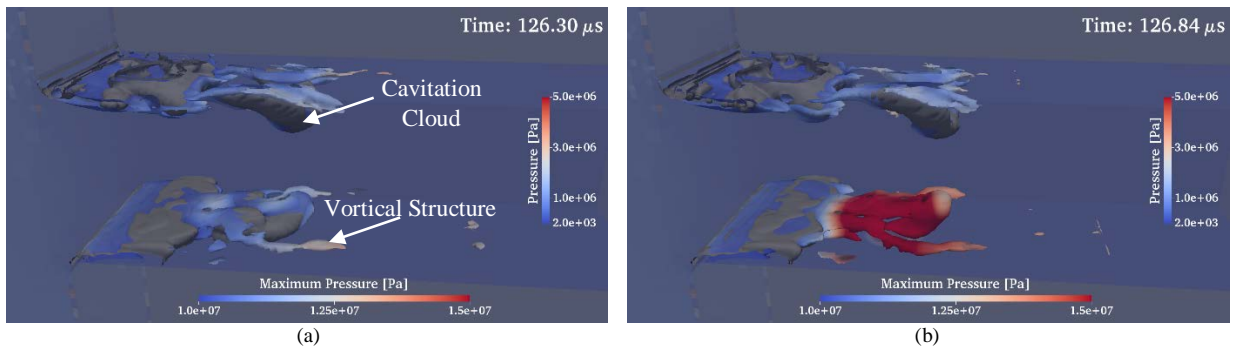

Figure 4. Predicted vortical structures for the pressure drop condition of 75 bar are visualized using iso-contours of vorticity magnitude of 5.0e6 Hz, and colored by the local pressure. Cavitation cloud structures, as previously shown in Figures 3(a) and (c), are overlaid in grey to highlight the vortex-cavitation cloud interaction leading up to the predicted cavitation erosion event.

**Summary**

In this work, a homogeneous relaxation model was employed within a large eddy simulation to predict cavitation phenomenon in the Winklhofer Throttle "U" geometry. Potential cavitation erosion events, identified using the maximum predicted pressure at the throttle boundary, were observed to be linked to the collapse of horseshoe cloud structures. Future investigations will examine the predicted vortical structures that are formed in the cavitation shedding and condensation processes, and further assess indicators that are capable of linking the cavitation erosion potential with the observed cavitation cloud structures.

**Acknowledgements**

*Corresponding Author, Gina Magnotti: gmagnotti@anl.gov

nonexclusive, irrevocable worldwide license in said article to reproduce, prepare derivative works, distribute copies to the public, and perform publicly and display publicly, by or on behalf of the Government.

The authors gratefully acknowledge the computing resources provided on Blues, a high-performance computing cluster operated by the Laboratory Computing Resource Center at Argonne National Laboratory, and Convergent Science Inc., for providing the Converge CFD software licenses.

*Corresponding Author, Gina Magnotti: gmagnotti@anl.gov